\documentstyle[letter]{ptptex}

\title{
{\bf $ SU(3)\times SU(2)\times U(1)$} Chiral Models from 
Intersecting D4-/D5-Branes
}

\author{

Hironobu {\sc Kataoka}$^1$\footnote{E-mail: kataoka@yukawa.kyoto-u.ac.jp} 
and 
Masafumi {\sc Shimojo}$^2$\footnote{E-mail: shimo0@ei.fukui-nct.ac.jp} 
}

\inst{
$^1$ Department of Physics, Hyogo University of Education,
Yashiro-cho, Hyogo 673-1494, Japan,
\\
$^{2}$Department of Electronics and Information Technology, 
Fukui National College of Technology, Sabae 916-8507, Japan
\\
}

\recdate{
December 28, 2001
}

\abst{
We clarify RR tadpole cancellation conditions for intersecting 
D4-/D5-branes. We determine all of the D4-brane models that have D=4 
three-generation chiral fermions with 
$SU(3)\times SU(2)\times U(1)^n$ symmetries. 
For the D5-brane case, we present a solution to the conditions 
that gives exactly the matter content of the standard model with 
$U(1)$ anomalies.
}

\begin{document}

\maketitle


Intersecting D$p$-branes are useful tools to obtain chiral 
fermions existing on their intersections.\cite{rf:AF}  
Although configurations of type IIB D6-branes wrapped on ${\bf T}^6$ 
give rise to identically the matter content of the standard 
model,$^{\ref{rf:IM})}$ the D6-brane models do not solve the 
fine-tuning problem caused by its breakdown of supersymmetry. 
Contrastingly, this difficulty is avoided by  
using D4-/D5-branes wrapped on ${\bf T}^{2d}\times {\bf R}^{6-2d}/Z_N$ 
through reduction of the string scale with a very large 
volume of the $Z_N$ orbifold. 
Intersecting D4-/D5-branes in both oriented and 
unoriented theories, we construct D=4 chiral models 
whose matter content transforms as $(3,2)_{\frac{1}{6}}$+
$(\overline{3},1)_{-\frac{2}{3}}$+ $(\overline{3},1)_{\frac{1}{3}}$ 
under $SU(3)\times SU(2)\times U(1)_Y$. 

We start with the definition of the spacetime geometry in the 
oriented theory. The compactified six dimensional space is 
the product of $d$ rectangular two-tori 
${\bf T}_I^2\ (1 \leq I\leq d\leq 2)$ and a ${\bf R}^{6-2d}/Z_N$ orbifold. 
The space is parameterized by the complex coordinates 
$Y_I=X_{2I+2}+iX_{2I+3}$ with radii $R_{2I+2}$ and $R_{2I+3}$ for $I=1,2,3$. 

Let us consider $K$ different stacks, each of which 
is a set of $N_a$ coincident D4-/D5-branes. They are labeled by the index 
$a$$(a=1,\cdots,K)$ and wrapped around a 1-cycle denoted by 
$(n_a^{(I)}, m_a^{(I)})$ on each of the $d$ two-tori, 
where $I=1$ for D4-branes and $I=1,2$ for D5-branes. 
When the $a$-th stack and the $b$-th stack intersect, 
we refer to their intersections as an $a$-$b$ sector. 
The intersection number of the $a$-$b$ sector is given by 
\begin{equation}
I_{ab}=\prod_{I=1}^dI_{ab}^{(I)}=
\prod_{I=1}^d(n_a^{(I)}m_b^{(I)}-n_b^{(I)}m_a^{(I)}).
\label{iab}
\end{equation}

The $Z_N$ action is realized by powers of the twist generator 
represented by a shift vector $v$ in the space of the $SO(6-2d)$ Cartan 
subalgebra. In the configurations with intersecting D4-branes, 
modular invariance conditions for one-loop amplitudes 
restrict the form of the twist vector 
$v=(v_1, v_2, v_3)$ to $(0,1,b_2)/N$, where $b_2$ is an odd integer. 
For the case of D5-branes, $v=(0,0,2/N)$. 

Since $N_a$ coincident D-branes have $U(N_a)$ gauge factors in 
their world-volume, open strings stretched between the $a$-th stack and 
the $b$-th stack give rise to states in the $a$-$b$ sector 
that belong to bifundamental representations of $U(N_a)\times U(N_b)$. 
The gauge degrees of freedom are 
expressed by the Chan-Paton (CP) factors as $|ij\rangle $, where $i$ and $j$ 
correspond to a D-brane in the $a$-th and the $b$-th stack, respectively. 
We express the embedding of the $Z_N$ action $\theta^k$  
using the unitary matrix $\gamma_k$.
We diagonalize $\gamma_{k}$ so that
\begin{equation}
\gamma_k={\rm diag}
(\zeta^{l_1}{\bf 1}_{N_{1}},
\zeta^{l_2}{\bf 1}_{N_{2}},
\cdots,\zeta^{l_K}{\bf 1}_{N_K}),
\end{equation}
where $\zeta=e^{2\pi i\frac{1}{N}}$ and $l_a$ are integers.
Then the $N_a$ D-branes in the $a$-th stack have the same CP phase, 
$\zeta^{l_a}$. 

The RR tadpole cancellation conditions$^{\ref{rf:AF})}$\cite{rf:blumen} are 
\begin{eqnarray}
\sum_{a=1}^KN_a\prod_{I=1}^dn_a^{(I)} =   
\sum_{a=1}^KN_a\prod_{I=1}^dm_a^{(I)} & = & 0, \label{rrtnor1}\\
\sum_{k=1}^{N-1}\left(\prod_{J=d+1}^3|{\rm sin}\pi k v_J|\right)
\left(\sum_{a=1}^K (\prod_{I=1}^dn_a^{(I)}){\rm Tr}_a\gamma_k\right)^2 
& = & 0,\label{rrtnor2} \\
\sum_{k=1}^{N-1}\left(\prod_{J=d+1}^3|{\rm sin}\pi k v_J|\right)
\left(\sum_{a=1}^K (\prod_{I=1}^dm_a^{(I)}){\rm Tr}_a\gamma_k\right)^2 
& = & 0, \label{rrtnor3}
\end{eqnarray}
where ${\rm Tr}_a$ represents the trace over the CP factor on 
D-branes that belong to the $a$-th stack.
For each shift vector $v$ of $Z_N$, we can obtain solutions 
for 1-cycles $(n_a^{(I)},m_a^{(I)})$ of the $a$-th stack and CP phases 
$\zeta^{l_a}$. The solution produces an open string spectrum 
that is invariant under the $Z_N$ action.\cite{rf:doug}

For the D4-brane case, left-handed 
fermions in the $a$-$b$ sector are given in bifundamental 
representations under $U(N_a)\times U(N_b)$ as 
$^{\ref{rf:AF})}$
\begin{equation}
I_{ab}\{(N_a^i,\overline{N_b}^{i-\frac{1-b_2}{2}})
+(N_a^i,\overline{N_b}^{i+\frac{1-b_2}{2}})
+(\overline{N_a}^i,N_b^{i-\frac{1+b_2}{2}})
+(\overline{N_a}^i,N_b^{i+\frac{1+b_2}{2}})\},
\end{equation}
where the symbol $N_a^i$ expresses the meaning that this is the 
representation $N_a$ of $U(N_a)$ and the CP phase 
of the $a$-$th$ stack is $\zeta ^i$.
A negative value of $I_{ab}$ indicates a positive multiplicity of 
the field that transforms as the complex conjugate 
of each bifundamental representation. 
To obtain the three (3,2) representations, we require $N_1=3$ and $N_2=2$ 
stacks and the intersection number $I_{12}=3$. 
The two types of three-generation right-handed quarks require  
$N_3=1$ and $N_4=1$ stacks and intersection numbers 
$I_{13}=\pm 3$ and $I_{14}=\pm 3$.  One more stack $N_5=1$ is 
necessary for the consistency conditions (\ref{rrtnor1})-- 
(\ref{rrtnor3}) to be satisfied. We have investigated all of 
these configurations and found two solutions, given  
in Table I. 
\begin{table}
\caption{D4-brane 1-cycles and CP phases giving rise to three generations of 
quarks.}
\begin{center}
\begin{tabular}{|l|c|c|c|c|c|c|c|}
\hline 
\hline
 & \raisebox{-2.5ex}{$Z_N$} & & \multicolumn{5}{c|}{cycle} \\
 & & & \multicolumn{5}{c|}{CP phase} \\  
\cline{4-8}
\raisebox{3.5ex}{number} & \raisebox{2.5ex}{orbifold} & \raisebox{3.5ex}{$b_2$} &
$N_1=3$ & $N_2=2$ & $N_3=1$ & $N_4=1$ & $N_5=1$ \\
\hline
 & & & $(1,0)$ & $(n_2,3)$ & $(n_3,-3)$ & $(-2n_2-n_3,-3)$ & $(-3,0)$ \\
 \raisebox{1.5ex}{D4-1} & \raisebox{1.5ex}{$N\geq 3$} 
 & \raisebox{1.5ex}{$b_2\neq -1$} &  1
 & $\zeta ^\frac{b_2-1}{2}$ & $\zeta^{\frac{b_2-1}{2}}$ &
 $\zeta^{\frac{b_2-1}{2}}$ & $1$ \\
\hline
 & & & $(1,0)$ & $(n_2,3)$ & $(n_3,-3)$ & $(-2n_2-n_3,-3)$ & $(-3,0)$ \\
\raisebox{1.5ex}{D4-2} & \raisebox{1.5ex}{$N\geq 3$} 
& \raisebox{1.5ex}{$b_2=-1$} &  1
 & $\zeta ^\frac{b_2-1}{2}$ & $\zeta^{\frac{b_2-1}{2}}$ &
 $\zeta^{\frac{b_2-1}{2}}$ & $1$ \\
 \hline
\end{tabular}
\vspace{10pt}
\end{center}
\end{table}

Since the gauge symmetry $U(1)$ of the stack $N_a=1$ with 1-cycle $(n_a,0)$ 
extends to $U(1)^{|n_a|}$,\cite{rf:AF}
the gauge symmetry of the models is $U(3)\times U(2)\times 
U(1)^5=SU(3)\times SU(2)\times U(1)^7$. The fermion spectrum 
under the symmetry given by D4-1 in  
Table I is as follows:   
\begin{equation}
\begin{array}{l}
3(3,2)_{(1,-1,0,0,0^3)}+3(\bar{3},1)_{(-1,0,1,0,0^3)}
+3(\bar{3},1)_{(-1,0,0,1,0^3)}  \\
+3(1,2)_{(0,1,0,0,\underline{-1,0,0})}
+3(1,1)_{(0,0,-1,0,\underline{1,0,0})}+3(1,1)_{(0,0,0,-1,\underline{1,0,0})},
\end{array}
\label{spd4-1}
\end{equation}
where the underlines indicate permutation of indices. 
Computing the mixed anomalies which need to be cancelled by the 
Green-Schwartz mechanism, we find six non-anomalous $U(1)$ linear 
combinations. They include the suitable hypercharge
\begin{equation}
Q_Y=-\frac{1}{6}Q_2+\frac{1}{3}Q_3-\frac{2}{3}Q_4+\frac{1}{3}(Q_5^{(1)}+Q_5^{(2)})
-\frac{2}{3}Q_5^{(3)},
\label{hyd41}
\end{equation} 
where $Q_a$ is the generator of the $a$-th $U(1)$. 
With this hypercharge, the fermion spectrum is 
\begin{equation}
\begin{array}{l}
3(3,2)_{\frac{1}{6}}+3(\overline{3},1)_{\frac{1}{3}}+
3(\overline{3},1)_{-\frac{2}{3}}+6(1,2)_{-\frac{1}{2}}
+3(1,2)_{\frac{1}{2}}\\
\ \ \ +6(1,1)_1+9(1,1)_0+3(1,1)_{-1}.
\end{array}
\label{spd4-1y}
\end{equation} 
When we set $n_2+n_3=1$, D4-2 in Table I gives the following fermion spectrum:
\begin{equation}
\begin{array}{l}
3(3,2)_{(1,-1,0,0,0^3)}+3(\bar{3},1)_{(-1,0,1,0,0^3)}+
3(\bar{3},1)_{(-1,0,0,1,0^3)} \\
+3(1,2)_{(0,1,0,0,\underline{-1,0,0})}
+3(1,1)_{(0,0,-1,0,\underline{1,0,0})}+3(1,1)_{(0,0,0,-1,\underline{1,0,0})}
\\
+6(1,2)_{(0,1,-1,0,0^3)}+6(1,2)_{(0,-1,0,1,0^3)}+12(1,1)_{(0,0,1,-1,0^3)}. \\
\end{array}
\label{spd4-2}
\end{equation}
We choose the hypercharge from 
anomaly-free realizations of $U(1)$ as 
\begin{equation}
Q_Y=\frac{1}{3}Q_1+\frac{1}{6}Q_2+\frac{2}{3}Q_3-\frac{1}{3}Q_4
-\frac{1}{3}(Q_5^{(1)}+Q_5^{(2)}+Q_5^{(3)}).
\label{hyd42}
\end{equation} 
With this hypercharge, the spectrum (\ref{spd4-2}) become  
\begin{equation}
\begin{array}{l}
3(3,2)_{\frac{1}{6}}+3(\overline{3},1)_{\frac{1}{3}}+3(\overline{3},1)_{-\frac{2}{3}}
+12(1,2)_{-\frac{1}{2}}+9(1,2)_{\frac{1}{2}}\\
\ \ \ +12(1,1)_1+9(1,1)_0+9(1,1)_{-1}.
\end{array}
\label{spd4-2y}
\end{equation}
When we set $n_2+n_3=-1$, we get the same spectrum (\ref{spd4-2y}) 
by exchanging $Q_3$ and $Q_4$ in the hypercharge (\ref{hyd42}).
For the case $n_2+n_3\neq \pm 1$, we have not been able to obtain the 
solution with the correct hypercharge $Q_Y$.

When the $a$-th stack and $b$-th stack have the same CP phase, 
there are $I_{ab}$ tachyon fields in the $a$-$b$ sector.\cite{rf:AF} 
In these models, the 2-3 and 2-4 sectors give rise to 
tachyon fields that transform as the $SU(2)$ doublet. 
This indicates that the configurations of branes that have $U(2)$ and $U(1)$ 
symmetries are unstable. This instability suggests a stringy Higgs mechanism of 
electroweak symmetry breaking.$^{\ref{rf:AF})}$    

For the D5-brane case, the fact that there are 
many wrapping numbers $(n_a^{(I)}$, $m_a^{(I)})$ with $(I=1,2)$ makes it 
difficult to obtain general solutions.  
For this reason, in this paper, we content ourselves with a single example. 
We will return to this problem 
at the end of the investigation of the unoriented theory.

We now discuss the unoriented theory. We introduce an orientifold group 
written as $Z_N+\Omega RZ_N$, where $\Omega$ is the world sheet parity. 
$R$ transforms $Y_I$ to $\overline{Y_I}$ for 
$1\leq I\leq d$ and $Y_I$ to $-Y_I$ for 
$d+1\leq I \leq 3$. The space-time is expressed as 
$$\frac{{\cal M}_4\times \prod_{I=1}^d{\bf T_I^2} 
\times (\prod_{I=1+d}^3 {\bf T_I^2}/Z_N)}{\Omega R}.$$
We express the $\Omega R$ action on the CP factors using 
unitary matrices $\gamma_{\Omega R}$. 
Since $\Omega R \theta^k (\Omega R)^{-1}$ $=\theta ^{2k}$,
we have
\begin{equation}
\gamma _{\Omega R k}=\gamma _k\gamma_{\Omega R}=
\pm\gamma_{2k}\gamma_{\Omega R k}^{T}.
\label{g2kgor}
\end{equation}
This leads to 
$\gamma_{\Omega R} =  \pm \gamma_{\Omega R}^T$.
In the following equations, the upper and lower sign correspond 
to the symmetric and antisymmetric $\gamma_{\Omega R}$, respectively.
 
In addition to the $a$-$b$ sector, there are intersections of 
the $a$-th stack and mirror branes of the $b$-th stack, which we refer 
to as $a$-$b*$ sector. The intersection number  
$I_{ab*}$ is obtained by changing  
$m_b^{(I)}$ to $-m_b^{(I)}$ in the expression (\ref{iab}).

The $Z_N$ action with even $N$ contains an order 2 element, so that
the orientifold group contains an element 
that does not vary $Y_I(I\geq d+1)$. 
Then there will be D8-branes in the type IIA D4-brane theory. 
Since the concept of intersecting D-branes involves use of the same 
dimensional D-branes, we restrict ourselves to the case that 
the order $N$ of $Z_N$ is odd. 

Tadpole divergences come from Klein bottle and M\"obius strip 
amplitudes as well as cylinder amplitudes.\cite{rf:blumen}
The RR tadpole cancellation conditions common to configurations 
with intersecting D4-branes and configurations with D5-branes 
are given by 
\begin{eqnarray}
\sum_{a=1}^KN_a\prod_{I=1}^dn_a^{(I)} & = &  \pm 16, \label{tcNn} \\  
\sum_{a=1}^KN_a\prod_{I=1}^dm_a^{(I)} & = & 0 \label{nmama}, \\
\sum_{a=1}^K\left(\prod_{I=1}^dm_a^{(I)}\right){\rm Tr}_a\gamma_k & = & 0 
\ \ \ {\rm for}\ k=1,\cdots ,N-1.
\label{nmatr}
\end{eqnarray}
There is one more condition on the product of 
the cycles $n_a^{(I)}$ and the trace of $\gamma_k$. 
When we consider a $Z_3$ orbifold, it is given by 
\begin{equation}
\sum_{a=1}^Kn_a^{(1)}{\rm Tr}_a\gamma_k=\pm 4 \ \ \ {\rm for}\ k=1,2,
\label{ntrg}
\end{equation}
for the D4-brane case and
\begin{equation}
\sum_{a=1}^Kn_a^{(1)}n_a^{(2)}{\rm Tr}_a\gamma_k=\mp 8 
\ \ \ {\rm for}\ k=1,2,
\label{nntrg}
\end{equation}
for the D5-brane case. For other $Z_N$ models with odd $N$, the condition 
is expressed by a linear summation of $\prod_{J=d+1}^3{\rm sin}\pi kv_J$ 
and $\prod_{J=d+1}^3{\rm sin}2\pi kv_J$ over $k=1,\cdots, N-1$ similar to 
(\ref{rrtnor2}), and we could not find any solution to 
1-cycles and $\gamma$ matrices. 

In order to satisfy the conditions (\ref{tcNn}) and 
(\ref{ntrg}) for unoriented D4-brane configurations 
giving a standard-like model, 
we must introduce many $U(1)$ stacks or 
an $N_2=2$ stack that has a 1-cycle $n_2^{(1)}\neq 1$.
Some of these $U(1)$ stacks and the $N_1=3$ stack always have 
intersections that lead to non-standard 
$(3,1)$ or $(\overline{3},1)$
representations under $SU(3)\times SU(2)$. 
The 1-cycles $n_2^{(1)}\neq 1$ in the 
unoriented theory produce fields that transform as a three-dimensional 
representation under $U(2)$.\cite{rf:IM} Thus we were not able to 
obtain the matter content of the standard model.

For models with intersecting D5-branes,
left-handed fermions in the $a$-$b$ sector and $a$-$b*$ sector
are given by$^{\ref{rf:AF})}$ $^{\ref{rf:IM})}$
\begin{equation} 
I_{ab}((N_a^i,\overline{N_b}^{i+1})+(\overline{N_a}^i,N_b^{i-1})),
\end{equation}
\begin{equation}
I_{ab*}((N_a^i,N_b^{-i-1})+(\overline{N_a}^i,\overline{N_b}^{-i+1})),
\label{d5iab*}
\end{equation}
under $U(N_a)\times U(N_b)$.

The bifundamental representations of (\ref{d5iab*}) for the $a$-$a*$ 
sector change to $N_a^i\wedge N_a^{-i-1}$ and $\overline{N_a}^i\wedge
\overline{N_a}^{-i+1}$ under $U(N_a)$.
On-orientifold intersections of the $a$-$a*$ sector
give $4m_a^{(1)}m_a^{(2)}$ fermions in the representation of 
either an antisymmetric or symmetric tensor, depending on whether 
$\gamma_{\Omega R}$ is antisymmetric or symmetric. Off-orientifold intersections of the 
$a$-$a*$ sector produce $2m_a^{(1)}m_a^{(2)}(n_a^{(1)}n_a^{(2)}-1)$ 
symmetric and antisymmetric representations. We require that any 
brane stack satisfy $m_a^{(1)}m_a^{(2)}=0$ to avoid the appearance of 
exotic quantum numbers and to satisfy the RR tadpole cancellation 
conditions  (\ref{nmama}) and (\ref{nmatr}). Adding D5-branes with 
$m^{(1)}=m^{(2)}=0$ will satisfy the conditions (\ref{tcNn}) and 
(\ref{nntrg}) without modifying the fermionic matter content. 

Since $I_{12}=-I_{12*}$ under 1-cycles for which $m_a^{(1)}m_a^{(2)}=0$, 
we must set $I_{12}=\pm 3$ and choose CP factors yielding no fermions 
in the 1-$2*$ sector to get $3(3,2)$ representations of $U(3)\times U(2)$. 
This configuration makes 
it impossible to obtain just the standard model spectrum without 
$U(1)$-$U(1)$-$U(1)$ anomalies.$^{\ref{rf:IM})}$
An example is given in Table II for the case in which the orbifold group is $Z_3$.
\begin{table}
\caption{Example of D5-brane 1-cycles and CP phases 
giving rise to three generations of quarks. 
The orbifold group is $Z_3$ and the parameter $\zeta$ is $e^{2\pi i/3 }$.}
\begin{center}
\begin{tabular}{|c|c|c|c|}
\hline 
\hline 
$N_a$ & $(n_a^{(1)},m_a^{(1)})$ & $(n_a^{(2)},m_a^{(2)})$ & CP phase \\
\hline
$N_1=3$ & $(n_1^{(1)},0)$ & $(n_1^{(2)},3/n_1^{(1)})$ & $\zeta^2$\\
\hline
$N_2=2$ & $(n_2^{(1)},m_2^{(1)})$ & $(1/m_2^{(1)},0)$ & $\zeta $ \\
\hline
$N_3=1$ & $(n_3^{(1)},m_3^{(1)})$ & $(-1/m_3^{(1)},0)$ & $\zeta $\\
\hline 
$N_4=1$ & $(n_4^{(1)},m_4^{(1)})$ & $(-1/m_4^{(1)},0)$ & $\zeta $ \\
\hline
$N_5=1$ & $(n_5^{(1)},0)$ & $(n_5^{(2)}, -3/n_5^{(1)})$ & $\zeta^2$\\
\hline
\end{tabular}
\end{center}
\end{table}

The spectrum under the gauge symmetry 
$SU(3)\times SU(2)\times U(1)^5$ 
is given by 
\begin{equation}
\begin{array}{c}
3(3,2)_{(1,-1,0,0,0)}+3(\overline{3},1)_{(-1,0,1,0,0)} 
+3(\overline{3},1)_{(-1,0,0,1,0)} \\ 
+3(1,2)_{(0,1,0,0,-1)} +3(1,1)_{(0,0,-1,0,1)}+3(1,1)_{(0,0,0,-1,1)}.
\end{array} \label{spd5}
\end{equation}
Although the spectrum has a $U(1)$-$U(1)$-$U(1)$ anomaly and 
$G^2$-$U(1)$ mixed anomalies that must be removed by some mechanism, 
two anomaly-free $U(1)$\ linear combinations exist. We can define 
the hypercharge as 
\begin{equation}
Y=\frac{1}{6}Q_1+\frac{1}{2}(Q_3-Q_4+Q_5).
\end{equation} and obtain identically the standard model spectrum under 
$SU(3)\times SU(2)\times U(1)_Y$ gauge symmetry 
with three generations of right-handed neutrino. 

We now consider the construction of D5-branes in the oriented theory.
In this case, any matter fermions (\ref{spd5}) for the unoriented theory are 
obtained from the $a$-$b$ sector, not from the $a$-$b*$ sector.
Adding an appropriate number of $U(1)$ branes to the stacks in Table II, 
we obtain D5-branes for the oriented theory that satisfy the 
conditions on the 1-cycles and CP phases (\ref{rrtnor1})--(\ref{rrtnor3}).   


Configurations with intersecting D-branes cause the breakdown of 
supersymmetry. Then the string scale must be close to the weak scale 
to avoid the fine-tuning problem. The four dimensional 
Plank mass $M_p$ and string scale $M_s$ are related as 
$M_p\approx 4\pi^2M_s^4\sqrt{V_TV_{Z_N}}/\lambda _{II}$, 
where $\lambda _{II}$ is the Type II string coupling, $V_T$ is the volume of 
the tori, and $V_Z$ is the volume of the orbifold. 
While a very large value of $V_T$ leads to 
small Yukawa and gauge couplings, we can give a very large volume 
to the transverse $Z_N$ orbifold. For models of D5-branes, 
for example, we can set $M_s \approx 1 \sim 10$ TeV and 
$V_T\sim 1/M_s^2$. In order to obtain very large value of the Plank mass, 
we should choose $\sqrt{V_{Z_N}}\approx 10^9$ -- $10^{11}$ (GeV)$^{-1}$, 
i.e., $10^{-2}$ -- $10^{-4}$ cm. 

In this paper, we have not investigated all of the 
standard-like spectra that accompany intersecting D5-branes. 
In particular, unoriented models may be obtained from 
1-cycles for which $m_a^{(1)}m_a^{(2)}\neq 0$, $n_a^{(1)}n_a^{(2)}=1$. 
In such models, the $a$-$a*$ sector may produce the antisymmetric tensors of 
$N_a\wedge N_a$ under $U(N_a)$ for $N_a=2,3$ without symmetric tensors 
that are not present in the matter content of the standard model. 



\end{document}